\documentclass{PoS}

\title{Jet masses for high $p_T$ QCD jets at all orders.}

\ShortTitle{QCD jet masses}

\author{\speaker{Mrinal Dasgupta}\thanks{The author, unable to attend the conference in person due to the volcanic ash cloud, would like to thank A.~Banfi 
for giving this talk on his behalf at DIS 2010. }\\
        University of Manchester\\
        E-mail: \email{Mrinal.Dasgupta@manchester.ac.uk}}

\abstract{The study of the shape and sub-structure of high $p_T$ jets produced in hadron collisions is becoming an increasingly important component of LHC phenomenology in the context of new particle discoveries. We study here the state of the art for masses of QCD jets in perturbation theory to all orders, up to next-to--leading logarithmic accuracy, with inclusion of non-global logarithms. We also study the precise role of the jet algorithm used to construct the jets and its impact on resummed predictions. Such analytical predictions, where 
available, offer an alternative to leading logarithmic predictions from Monte Carlo event generators and could, after fixed-order matching and accounting for non-perturbative effects, be used for direct comparison to early LHC data.}

\FullConference{XVIII International Workshop on Deep-Inelastic Scattering and Related Subjects, DIS 2010\\
		April 19-23, 2010\\
		Firenze, Italy}

\begin{document}

\section{Introduction}
Studies of jet-shape variables and in particular jet masses have long been a 
standard part of the program of testing perturbative QCD. The infrared and collinear safety of such observables has rendered them calculable by perturbative 
methods both in fixed-order perturbation theory as well as amenable to 
all-order resummed treatments in special regions of phase space enhanced by 
large logarithms \cite{CTTW}.   

In the context of the LHC, jet-shape and sub-structure techniques \cite{Seymour} are also rapidly gaining impetus for example as a means of discovering new particles that are highly boosted and hence whose decay products may end up in a single jet. As examples of recent sub-structure methods in this context one 
can refer to the development of jet filtering \cite{BSR} and pruning techniques \cite{Ellis} amongst a variety of tools being developed to reduce QCD jet 
backgrounds as well as to enable better reconstruction of mass-peaks for new heavy particles.

In this paper we concentrate on what is known about jet shapes and in particular jet masses for QCD jets in all orders resummed perturbation theory. At the 
LHC one may be interested for instance in the mass or shape of one or more jets produced in high $p_T$ jet events of a given jet multiplicity. Recently it was suggested to study such observables with the aim of resumming logarithms in the shape variables for certain jets in multi-jet events while leaving the shape of other jets unmeasured \cite{Ellis1,Ellis2}. Resummed predictions were 
provided aiming at next-to--leading or single-logarithmic accuracy in the shape variable distributions, within the context of soft-collinear effective theory (SCET). However as was emphasised in Ref.~\cite{BDKM} single logarithmic accuracy for such observables necessitates a study and inclusion of non-global logarithms \cite{NG1,NG2}. Below we shall examine this issue and comment on the size of the non-global effect as well as looking at the role of the jet algorithm 
which is also significant to single-logarithmic accuracy.

\section{Individual jet masses in multijet events and non-global logarithms}
We can consider either the jet-mass in single-jet inclusive studies or focus on events of fixed multiplicity where we may pick a subset of all jets for study, as was the case in Refs.~\cite{Ellis1,Ellis2}. 

To illustrate our points here we examine a simple situation involving high $p_T$ dijet production where we measure the mass of one of the jets while leaving the other unmeasured. We can take the jets to be produced at zero rapidity wrt 
the beam without changing our conclusions. We shall also consider the limit of narrow well-separated jets in the sense that the jet radius $R$ can be taken to be small compared to the interjet separation $\Delta_{ij} \equiv 1-\cos\theta_{ij}$. We shall thus neglect terms of relative order $R^2/\Delta_{ij}$. In 
this limit we find that owing to QCD coherence a rather simple picture of jet evolution emerges \cite{BDKM}. In particular initial state radiation only contributes terms to the jet mass that vanish in the $R \to 0$ limit and hence can be ignored in our approximation. We shall carry out fixed-order calculations below, in the soft approximation, both at LO and NLO accuracy and from these calculations infer the form of our resummed results.

\section{LO estimate in soft approximation}
A simple eikonal leading-order calculation for the cross-section for 
normalised jet mass $4 M^2/Q^2$ to be below some value $\rho$ \cite{BDKM} produces
\begin{equation}
\label{eq:nlores}
\Sigma_{\mathrm{LO}} \sim - \frac{C_F\alpha_s}{\pi} \int_{\rho Q/2R^2}^{Q/2} \frac{d\omega}{\omega} \ln \left (2\frac{\omega R^2}{Q \rho} \right) \\
=-\frac{C_F \alpha_s}{2\pi} \ln^2 \frac{R^2}{\rho} \Theta \left(R^2-\rho \right),
\end{equation}
where $Q$ is the hard scale of the process, in this  case the jet $p_T$. 
The above result holds no surprises -- it is the familiar double-log estimate for jet-mass, with $R$ being the jet radius. Such double logarithmic terms 
exponentiate and produce the Sudakov peak in the mass distribution. The calculation can be easily extended to account for hard emission collinear to the jet, which is a relevant source of single logarithms. However for the jet mass distribution at single-log accuracy there is more to the story than exponentiation of a single gluon. We report on this in the next section but note that at the level of terms in the conventional Sudakov exponent there is no involvement of the details of the algorithm except the radius $R$ -- all algorithms are identical at this level.

\section{NLO estimates and non-global logarithms}
Beyond leading order there is the issue of the soft gluon emission pattern as well as its interplay with the jet algorithm both of which lead to the appearance of non-trivial effects. A simplified emission pattern where soft gluons are 
coupled directly to the hard emitting ensemble is sufficient at the Sudakov level. However as was shown in Refs.~\cite{NG1,NG2} for non-global observables that are affected by soft radiation in a delimited phase space region such as the interior of a jet the pattern of real-virtual cancellations that ensure the 
exponentiation of a single-gluon is spoiled. One ends up needing to consider multiple soft gluon emission from an arbitrarily complex ensemble involving not just the hard partons but also all subsequent soft emissions until one hits the veto scale (the jet mass). This complication starts at the two-gluon level with the soft correlated two gluon emission term which for the quark dijets we assume here, has a $C_F C_A$ colour factor. 

Moreover from the two gluon level onwards the details of the jet algorithm also become important due to soft gluon clustering effects. One algorithm where these effects can be ignored is the anti-$k_t$ algorithm \cite{AKT} in which soft gluons cluster to the hard jets long before their self-clustering can take 
place, leading to circular jets. In this algorithm it proves possible, at least in the large $N_c$ limit, to address relatively easily the complex dynamics leading to non-global logarithms. Carrying out the calculation for the emission of two soft gluons and focussing on the $C_F C_A$ piece we obtain
\begin{equation}
\label{eq:ngrho}
S_2 = -C_F C_A \frac{\pi^2}{3}\left (\frac{\alpha_s}{2\pi}\right)^2 
\ln^2\frac{2 E_0 R^2}{\rho Q} \Theta \left(\frac{2 E_0 R^2}{Q}-\rho \right).
\end{equation}

The above equation needs some explanation. Firstly we have introduced a parameter $E_0$ analogous to the parameter $\Lambda$ in Refs.~\cite{Ellis1,Ellis2}, which corresponds to the maximum energy flowing outside the hard high $p_T$ jets. Limiting $E_0$ thus corresponds to restricting the hard jet multiplicity while for the inclusive jet mass case $E_0$ should be understood to be of the order of the hard scale i.e typical jet $p_T$. Here we shall focus on the inclusive 
jet mass case $E_0 \sim p_T$ while in Ref.~\cite{BDKM} we also dealt with resummation of logarithms in $p_T/E_0$. We note that the above equation is correct with neglect of additional terms suppressed by powers of $R$ which are beyond our approximation. With this approximation in place the coefficient $\pi^2/3$ emerges and is the {\emph{same coefficient}} as the corresponding one that appears for the hemisphere jet masses. This fact is not a coincidence but a result of the dominance of the collinear singularity that appears due to $1/(k_1.k_2)$ terms in the squared matrix element for the emission of soft gluons $k_1$ and $k_2$. The non-global logarithms come essentially  from the edge of the jet and have an additional analytic dependence on the jet radius $R$ which can be ignored in a small $R$ approximation. 

In the case of the hemisphere jet-mass studied in Ref.~\cite{NG1} non-global logs come from a large-angle boundary i.e that between the hemispheres. In the present case this boundary is replaced by the circular jet boundary but the coefficients and resummation of non-global terms will be the same as for the hemisphere case up to terms that vanish with $R$. The resummed result for the non-global piece, obtained numerically and in the large $N_c$ limit, can then be essentially taken from Ref.~\cite{NG1} and reads 
\begin{equation}
S(t) = \exp \left (-C_F C_A \frac{\pi^2}{3} t^2 \left(\frac{1+(at)^2}{1+(bt)^c} \right)\right)
\end{equation}
with $t$ being the single-logarithmic evolution variable $t=\frac{1}{2\pi} \int_{e^-L}^{1} \frac{dx}{x} \alpha_s(xQ)$ and where $L = \ln \left(2 E_0 R^2\right)/(Q \rho)$, $a=0.85 C_A$, $b=0.86 C_A$, $c=1.33$. 
The non-global factor $S$ multiplies the result obtained by exponentiating the one-gluon result (\ref{eq:nlores}) after taking account of hard collinear emissions and the running coupling. 

\begin{figure}
\includegraphics[width=.4\textwidth]{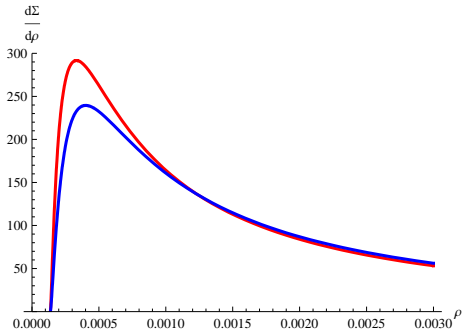}
\caption{Jet mass distribution with (blue curve) and without non-global logarithms (red curve) with $E_0= 60$ GeV and $R=0.4$.}
\label{fig1}
\end{figure}

The role of non-global logarithms in the jet mass distribution can be seen from Fig.~\ref{fig1}. The figure is for $Q=500$ GeV which translates here to a $250$ GeV jet $p_T$ with $R=0.4$ and $E_0 = 60$ GeV. There is roughly a 
twenty percent effect on peak height at these values of the parameters and in fact we find no significant variation in this number when $E_0$ is lowered due to the increasing importance of non-global logs in $p_T/E_0$ as reported in Ref.~\cite{BDKM}. The general situation where one measures the shapes of one or more jets in multi-jet production at high transverse momenta can be addressed by modifying the resummed result based on naive single-gluon exponentiation with a product of form factors from each jet, each of which has the form of $S(t)$ reported above. The above form is correct up to terms that vanish as powers of $R^2/\Delta_{ij}$ with $R$ the jet radius and $\Delta_{ij}$ the interjet 
separation. There is a clear physical reason for the emergence of the simple results reported here. In the small $R$ limit each jet evolves independently of others due to the dominance of collinear emissions -- the interesting physics (NLL contribution to jet masses) comes separately from the boundary of each jet which falls, at small $R$, within the collinear regime.

The results derived in this article are valid only in the anti-$k_t$ algorithm where soft gluons cluster to the hard jets independently of one another resulting in circular jets. This situation is modified when one moves to describing jets in other algorithms, which we shall discuss below.

\section{Other jet algorithms}
Let us now comment on the situation in jet algorithms other than the anti-$k_t$. Focussing on sequential recombination algorithms such as the Cambridge-Aachen (C-A) \cite{CA1,CA2} and the $k_t$ algorithm \cite{kt1,kt2} we encounter beyond leading double logarithms a much more complex situation. Appleby and Seymour \cite{AS} were the first to point out and numerically study the role of $k_t$ clustering on non-global logarithms (for the case of gaps between jets). Their conclusions were that the effect of $k_t$ clustering reduced the non-global component because a harder gluon $k_1$ is capable of clustering a softer emission $k_2$ in configurations that may otherwise have generated significant non-global effects. Similar conclusions will apply for the C-A algorithm. It was subsequently found that clustering also had an impact at NLL accuracy on real-virtual cancellations in the independent emission (as opposed to correlated emission) or global term \cite{BanDas}. While in Refs.~\cite{BanDas,BanDasDel} the clustering gave terms that vanished with $R$, in the present case owing to the collinear singularities relevant for jets at small $R$, the effect is independent of $R$. We find that for a quark jet the contribution of clustering to the $\rho$ distribution starts at $\mathcal{O}(\alpha_s)^2$ and the first such term reads
\begin{equation}
\frac{d}{d\rho} \Sigma_2^{\mathrm{cluster}} = -0.728C_F^2 \left( \frac{\alpha_s}{2\pi} \right)^2 \frac{1}{\rho} \ln \frac{1}{\rho}
\end{equation}
which is a relevant term for resummations aiming at single-log accuracy in the mass cross-section $\Sigma(\rho)$ and arises purely due to the clustering inherent in the algorithm. 
Likewise similar effects will be present for the C-A case. Single logarithmic terms generated by clustering were shown to be resummable for the gaps between jets case \cite{BanDasDel} but a similar calculation has not yet been carried out for the present case of jet masses.

\section{Conclusions}
In conclusion we point out that NLL resummed predictions can be made for quantities involving the shapes of jets produced in complex multi-jet events and using the anti-$k_t$ jet definition, which include non-global logarithms in the large $N_c$ limit. Up to correction terms varying as powers of the jet radius there is an independent non-global factor arising from the boundary of each measured jet. The product of such factors multiplies the exponentiated ``single-gluon'' result which is relatively straightforward to obtain. We do not expect neglected terms varying as powers of $R$ or those suppressed as $1/N_c^2$ to make a visible difference to our predictions even for $R$ as large as $0.7$ 
Phenomenological investigations can thus be carried out using the analytical results after matching to fixed-order to obtain the high-mass tail and assessing the role of non-perturbative effects.


\begin{thebibliography}{99}
 \bibitem{CTTW}
  S.~Catani, L.~Trentadue, G.~Turnock and B.~R.~Webber,
  ``Resummation of large logarithms in $e^+ e^-$ event shape distributions,''
  Nucl.\ Phys.\  B {\bf 407} (1993) 3.
\bibitem{Seymour}
  M.~H.~Seymour,
  ``Searches for new particles using cone and cluster jet algorithms: A
  Comparative study,''
  Z.\ Phys.\  C {\bf 62} (1994) 127.
\bibitem{BSR}
  J.~M.~Butterworth, A.~R.~Davison, M.~Rubin and G.~P.~Salam,
  ``Jet substructure as a new Higgs search channel at the LHC,''
  Phys.\ Rev.\ Lett.\  {\bf 100} (2008) 242001
  [arXiv:0802.2470 [hep-ph]].
 
\bibitem{Ellis}
 S.~D.~Ellis, C.~K.~Vermilion and J.~R.~Walsh,
 ``Techniques for improved heavy particle searches with jet substructure,''
 Phys.\ Rev.\  D {\bf 80} (2009) 051501
 [arXiv:0903.5081 [hep-ph]].

\bibitem{Ellis1}
  S.~D.~Ellis, A.~Hornig, C.~Lee, C.~K.~Vermilion and J.~R.~Walsh,
  ``Consistent Factorization of Jet Observables in Exclusive Multijet
  Cross-Sections,''
  arXiv:0912.0262 [hep-ph].
\bibitem{Ellis2}
  S.~D.~Ellis, A.~Hornig, C.~Lee, C.~K.~Vermilion and J.~R.~Walsh,
  ``Jet Shapes and Jet Algorithms in SCET,''
  arXiv:1001.0014 [hep-ph].
\bibitem{BDKM}
A.~Banfi, M.~Dasgupta, K.~ Khelifa-Kerfa and S.~ Marzani,
``High-$p_T$ jet shapes, non-global logarithms and jet algorithms,''
  arXiv:1004.3483 [hep-ph].
  

\bibitem{NG1}
  M.~Dasgupta and G.~P.~Salam,
  ``Resummation of non-global QCD observables,''
  Phys.\ Lett.\  B {\bf 512} (2001) 323
  [arXiv:hep-ph/0104277].
  \bibitem{NG2}
  M.~Dasgupta and G.~P.~Salam,
  ``Accounting for coherence in interjet E(t) flow: A case study,''
  JHEP {\bf 0203} (2002) 017
  [arXiv:hep-ph/0203009].
\bibitem{AKT}
  M.~Cacciari, G.~P.~Salam and G.~Soyez,
  ``The anti-$k_t$ jet clustering algorithm,''
  JHEP {\bf 0804} (2008) 063
  [arXiv:0802.1189 [hep-ph]].
\bibitem{CA1}
Y.~L.~Dokshitzer, G.D.~Leder, S.~Moretti and B.R.~Webber, 
``Better jet clustering algorithms'',
JHEP {\bf 9708} (1997) 01
[arXiv:hep-ph/9707323] 

\bibitem{CA2}
  M.~Wobisch and T.~Wengler,
  ``Hadronization corrections to jet cross sections in deep-inelastic
  scattering,''
  arXiv:hep-ph/9907280.



\bibitem{kt1}
  S.~Catani, Y.~L.~Dokshitzer, M.~H.~Seymour and B.~R.~Webber,
  ``Longitudinally invariant $K_t$ clustering algorithms for hadron hadron
  collisions,''
  Nucl.\ Phys.\  B {\bf 406} (1993) 187.

  \bibitem{kt2}
  S.~D.~Ellis and D.~E.~Soper,
  ``Successive combination jet algorithm for hadron collisions,''
  Phys.\ Rev.\  D {\bf 48} (1993) 3160
  [arXiv:hep-ph/9305266].
  
  
 \bibitem{AS}
  R.~B.~Appleby and M.~H.~Seymour,
  ``Non-global logarithms in inter-jet energy flow with $k_t$ clustering
  requirement,''
  JHEP {\bf 0212} (2002) 063
  [arXiv:hep-ph/0211426].




  \bibitem{BanDas}
  A.~Banfi and M.~Dasgupta,
  ``Problems in resumming interjet energy flows with $k_t$ clustering,''
  Phys.\ Lett.\  B {\bf 628} (2005) 49
  [arXiv:hep-ph/0508159].
  
\bibitem{BanDasDel}
  Y.~Delenda, R.~Appleby, M.~Dasgupta and A.~Banfi,
  ``On QCD resummation with $k_t$ clustering,''
  JHEP {\bf 0612} (2006) 044
  [arXiv:hep-ph/0610242].
\end{thebibliography}
\end{document}